\documentclass[12pt,preprint]{aastex}

\shorttitle{slipping reconnection and a triangle-flag flux rope}
\shortauthors{Li & Zhang}

\begin{document}

\title{Slipping Magnetic Reconnection Triggering a Solar Eruption of a Triangle-flag Flux Rope}

\author{Ting Li\altaffilmark{} \& Jun Zhang\altaffilmark{}}

\altaffiltext{}{Key Laboratory of Solar Activity, National
Astronomical Observatories, Chinese Academy of Sciences, Beijing
100012, China; [liting;zjun]@nao.cas.cn}

\begin{abstract}

We firstly report the simultaneous activities of a slipping motion
of flare loops and a slipping eruption of a flux rope in 131 {\AA}
and 94 {\AA} channels on 2014 February 02. The east hook-like flare
ribbon propagated slippingly at a speed of about 50 km s$^{-1}$,
which lasted about 40 min and extended by more than 100 Mm, but the
west flare ribbon moved in the opposite direction with a speed of 30
km s$^{-1}$. At the later phase of the flare activity, a ``bi-fan"
system of flare loops was well developed. The east footpoints of the
flux rope showed an apparent slipping motion along the hook of the
ribbon, simultaneously the fine structures of the flux rope
rose up rapidly at a speed of 130 km s$^{-1}$, much faster the whole
flux rope. We infer that the east footpoints of the flux rope are
successively heated by a slipping magnetic reconnection during the
flare, which results in the apparent slippage of the flux rope. The
slipping motion delineates a ``triangle-flag surface" of the flux
rope, implying that the topology of a flux rope is more complex than
anticipated.

\end{abstract}


\section{Introduction}

Magnetic reconnection causes the transition from magnetic energy to
kinetic energy of accelerated particles and thermal energy.
Energetic particles propagating along the reconnected magnetic
fields can impact the chromosphere, which results in the formation
of flare ribbons (Schmieder et al. 1996; Asai et al. 2004). Flare
ribbons are the footpoints of flare loops, which are formed due to
chromospheric evaporation (Milligan \& Dennis 2009; Ning et al.
2009). In the classical two-dimensional (2D) magnetic reconnection
model called the CSHKP model (Carmichael 1964; Hirayama 1974), the
magnetic field lines stretched by the erupting flux rope
successively reconnect at places where the magnetic connectivity is
discontinuous. A pair of flare ribbons are located in opposite
magnetic polarities and parallel to the magnetic polarity inversion
line (PIL).

However, magnetic reconnection can also take place at sites where
the magnetic connectivity is continuous but with a strong gradient,
i.e., quasi-separatrix layers (QSLs; Priest \& D{\'e}moulin 1995;
D{\'e}moulin et al. 1996). The flare ribbons are found to be along
or just near the intersection of QSLs, implying that reconnection
can occur in the absence of discontinuity (D{\'e}moulin et al. 1997;
Guo et al. 2013). Three-dimensional (3D) magnetohydrodynamic (MHD)
simulations showed that magnetic field lines passing through QSLs
can successively reconnect and exchange their connectivity with
neighboring field lines, which resulted in the apparent slipping
motion (Pontin et al. 2005; De Moortel \& Galsgaard 2006; Aulanier
et al. 2006).

Up to now, direct observations of slipping magnetic reconnection are
very rare. Aulanier et al. (2007) firstly reported fast
bidirectional motions of coronal loops, supportive of the existence
of slipping magnetic reconnection. Recently, Dud{\'{\i}}k et al.
(2014) presented the detailed observations that the flare loops
exhibited apparent slipping motion during an eruptive X1.4 flare. In
this work, the slipping motion of flare loops and the eruption of a
``triangle-flag" flux rope are simultaneously observed by the
\emph{Solar Dynamics Observatory} (\emph{SDO}; Pesnell et al. 2012).
Here, we focus on the evolution processes of the flare ribbons, the
slipping flare loops and the erupting flux rope.

\section{Observations and Data Analysis}

On 2014 February 02, a C8.9 flare occurred in NOAA AR 11967,
including two negative-polarity flare ribbons (NR1 and NR2 in Figure
1) and one positive-polarity ribbon (PR in Figure 1). The flare
loops connecting the hook-like NR1 and the PR exhibited an apparent
slipping motion with the west and east ends moving in opposite
directions. The slipping motion of the east footpoints of the flux
rope was also observed along the hook-like NR1.

The Atmospheric Imaging Assembly (AIA; Lemen et al. 2012) onboard
the \emph{SDO} takes full-disk images in 10 (E)UV channels at
1$\arcsec$.5 resolution and high cadence of 12 s. The flux ropes and
slipping flare loops could only be clearly observed in two
higher-temperature EUV channels of 131 {\AA} and 94 {\AA}. The 131
{\AA} channel best shows the flux rope and we focus on this channel
in this study. The high-cadence 304 {\AA} observations are used to
analyze the evolution of flare ribbons. Moreover, the observations
of 1600 {\AA} and 171 {\AA} are also used to show the
multi-wavelength appearance of this event. The 131 {\AA} channel
corresponds to a high temperature of about 11 MK (Fe VIII, Fe XXI)
and the channel of 304 {\AA} (He II) is at 0.05 MK (O'Dwyer et al.
2010; Boerner et al. 2012; Parenti et al. 2012). We also use the
full-disk line-of-sight magnetic field data from the Helioseismic
and Magnetic Imager (HMI; Scherrer et al. 2012) onboard \emph{SDO},
with a cadence of $\sim$ 45 s and a sampling of 0$\arcsec$.5
pixel$^{-1}$.

\section{Results}

\subsection{Evolution of Flare Ribbons and Slipping Motion of Flare Loops}

Before the flare started, the partial eruption of a filament was
initiated. Starting from 05:40 UT, the filament was initially
disturbed and associated with EUV brightening. Then partial filament
material rose up and seemed to be disconnected from its north end
(Figure 1(a); see Animation 304). At about 06:20 UT, the thread-like
structures overlying the filament gradually became clear and the
first flux rope (FR1 in Figures 2(a) and 3) appeared. When the flare
was initiated, another flux rope (FR2) gradually appeared at 07:14
UT (Figure 2(a)). The appearance of the FR2 was associated
with the partial activation of the filament, and the activated
filament material was embedded in the structures of the FR2. The AR
11967 where the flare occurred was complex and mainly consisted of
two couples of sunspots (Figure 2(b)).

Seen from AIA 304 {\AA} images, two flare ribbons started to appear
at two sides of the filament from about 07:05 UT (Figure 1(b)). The
GOES soft X-ray 1$-$8 {\AA} flux showed that the C8.9 flare
initiated at 07:17 UT, reached its peak at 07:44 UT and ended at
07:58 UT (Figure 4(a)). The PR moved to the south and its trajectory
was approximately parallel to the neutral line (Figures 1(b)-(h)).
Meanwhile, the NR1 propagated in the opposite direction and
delineated a hook-like morphology with a length of about 100 Mm.
Afterwards, the filament was disturbed again about 1.5 hr after its
activation at 05:40 UT. At about 07:10 UT, partial filament
material was activated and seemed to be disconnected from its north
end. The activated material moved along the strands of the filament
and erupted at last (see Animation 304). Subsequently another flare
ribbon (NR2 in Figures 1(c)-(h)) appeared at 07:18 UT. The HMI
line-of-sight magnetograms show that the PR is located in
positive-polarity fields nearby a sunspot of the AR (Figure 1(h)).
Before the appearance of the PR, the north magnetic structures of
the PR emerged and moved southward along the PR. The hook-like NR1
and NR2 lie at the negative-polarity region that corresponds to the
faculae of the AR (Figures 1(f) and (h)). In order to analyze the
propagation of the NR1 and PR, we obtain the stack plots (Figures
4(b)$-$(c)) along slices ``A$-$B" and ``C$-$D" (blue dotted lines in
Figure 1(b)) in 304 {\AA} images. The NR1 lasts about 40 min and the
average velocity of the NR1 is approximately 50 km s$^{-1}$. The PR
has a length of about 40 Mm, and propagates to the south with a
velocity of 30 km s$^{-1}$.

In the AIA 131 {\AA} and 94 {\AA} observations, the newly-formed
flare loops ``L1" and ``L2" connecting the NR1 and PR were observed
at about 07:19 UT (Figure 2(c)), when the flare started. ``L1" and
``L2" seemed to be crossed each other. Then flare loops ``L3," ``L4"
and ``L5" successively appeared and these flare loops formed a
``bi-fan" loop system (Figure 2(f)). The apparent slipping motion of
flare loops was clearly observed (Figures 2(c)-(f); Animation
131-loops). The west ends of the loops slipped to the southeast,
which developed into the PR. However, the east ends moved to the
north along the hook-like NR1. These slipping loops could not be
observed in low-temperature channels such as 171 {\AA} (Figure
1(g)).

\subsection{Slipping Eruption of the FR1}

The C8.9 flare seemed to trigger the slipping eruption of the FR1.
Since the appearance of the FR1, it started to expand and rise up
slowly at approximately 07:00 UT (Figures 3(a)-(b)). The east
footpoints of the FR1 showed an apparent slipping motion along the
hook of the NR1 (Figures 3(b)-(f); Animation 131-fluxropes). The
slipping motion of the FR1 lasted about 40 min, from about 07:06 UT
to 07:50 UT. Meanwhile, the formerly invisible fine structures of
the FR1 became visible in succession (solid curves in Figures
3(b)-(f)). These successively visible fine structures delineated a
``triangle-flag" topology of the FR1 (Figure 3(f)). However, the
west footpoints of the FR1 remained at the initial position. In
order to analyze the kinematic evolution of the FR1 in detail, we
obtain the stack plots (Figures 4(d)$-$(e)) along slices ``E$-$F"
and ``G$-$H" (Figure 3(a)). The initial rise velocity of the fine
structures of the FR1 is about 20 km s$^{-1}$ (Figure 4(d)). At
about 07:24 UT, the fine structures rose upward rapidly with a
velocity of 130 km s$^{-1}$. However, the whole FR1 rose up
very slowly at a speed of 5 km s$^{-1}$ from 07:18 UT to 07:38 UT
(Figure 4(e)). Afterwards, it was deflected towards the northeast
and the west end of the FR1 was broken away from the solar
photosphere due to its interaction with the north bundles of loops
in the north AR (Animation 131-fluxropes).

%

\section{Summary and Discussion}

We firstly report the simultaneous observations of the slipping
flare loops and the erupting FR1 by SDO/AIA on 2014 February 02 in
AR 11967. The ``triangle-flag surface" and the fine
structures, both along the erupting FR1, are also unprecedented as
compared to previous studies. The flare loops exhibited an apparent
slipping motion as seen in 131 {\AA} and 94 {\AA} channels. The west
and east footpoints of flare loops slipped in opposite directions,
resulting in the ``bi-fan" topology of flare loops. The east
hook-like flare ribbon at the faculae of the AR propagated at a
speed of about 50 km s$^{-1}$ and extended by more than 100 Mm. The
west flare ribbon moved in the opposite direction with a speed of 30
km s$^{-1}$, slightly smaller than the hook ribbon. Accompanying the
propagation of the hook-like ribbon, the east footpoints of the FR1
showed an apparent slipping motion along the hook of the ribbon. The
slipping footpoints were located nearby the position where the
hook-like ribbon propagated. The fine structures of the FR1
rose upward rapidly at a speed of about 130 km s$^{-1}$, much faster
than the whole FR1 (5 km s$^{-1}$). However, the west footpoints of
the FR1 were compact and stationary, anchoring in positive-polarity
magnetic flux concentration.

We considered the observed features of FR1 and FR2 as flux
ropes for two reasons. Firstly, the intertwined and twisted
structures are observed for FR1 and FR2. Gibson et al. (2004)
defined a flux rope as a set of EUV loops that collectively wind
around a central axial line. Secondly, FR1 and FR2 could only be
observed in high-temperature channels of AIA, such as 131 {\AA} and
94 {\AA}. In recent literatures (e.g., Cheng et al. 2011), they are
considered as high-temperature flux ropes. If FR1 and FR2 are other
features (e.g., potential loops), they usually could be observed in
both high- and low-temperature channels, including 171 {\AA}, 193
{\AA} and 131 {\AA} and 94 {\AA}. Thus, we are prone to name FR1 and
FR2 as flux ropes rather than other features.

The magnetic fields in QSLs are continuous and hence the change in
footpoint connectivity of the fields occurs at finite speed
(Aulanier et al. 2007). The speed of the slipping motion is the
exchange rate of connectivity of magnetic fields. According to the
speed of slipping motion, Aulanier et al. (2006) gave two
definitions of slip-running and slipping reconnection regimes that
respectively correspond to super- and sub-Alfv\'{e}nic field line
fast slippage. In our observations, the slipping speeds were 30 and
50 km s$^{-1}$. They are comparable to the slipping speeds of
coronal loops in Aulanier et al. (2007) and flare loops in
Dud{\'{\i}}k et al. (2014). These speeds are sub-Alfv\'{e}nic
speeds, and satisfy the slipping reconnection regime.

Our observations of an eruptive flare have revealed several typical
3D physical process, including: the hook-like flare ribbon and the
slippage of post-flare loops with two ends along opposite
directions. However, the CSHKP model could not completely account
for these observed properties. In the classical CSHKP model, the
flare ribbons are almost parallel to the PIL and flare loops have
similar shapes that are approximately perpendicular to the PIL.
Recently, 3D extensions to the CSHKP model have been carried out by
Aulanier et al. (2012) and Janvier et al. (2013). They have
simulated the 3D MHD evolution process of a torus-unstable erupting
flux rope during an eruptive flare, and the slipping-running
reconnection of field lines is well reproduced in their simulations.
Magnetic field lines undergo successive reconnections as they cross
the QSLs and the continuous rearrangements of field lines along the
QSLs generate the apparent slipping motion of field line footpoints.
Their ``standard solar flare model in 3D" matches our observations
in general.

Recently, direct observations of flux ropes with the \emph{SDO} data
have been reported by many authors (Li \& Zhang 2013a, 2013b, 2013c;
Yang et al. 2014). However, the slipping motion of the footpoints of
a flux rope is firstly observed in our work. We suggest that the
slippage of the footpoints of the FR1 are triggered by the slipping
magnetic reconnection during the flare process. For the footpoints
of the FR1 are close to the footpoints of reconnected magnetic
fields, the slipping magnetic reconnection successively heats the
footpoints of the FR1 and the pre-existing invisible structures
becomes visible. The successively visible structures form a surface
of a ``triangle flag" and indicates the FR1 has a ``triangle-flag"
topology. The fine structures of the FR1 rise upwards much faster
than the whole flux rope does, and seem to be continuously fed from
below while their easternmost footpoints keep slipping along the
flare ribbon. This is a clear and a first ever-observational
signature of the decoupling between the slowly moving plasma and the
fast slipping loops.

We suggest that the ``triangle-flag surface" may correspond
to one-half of the coronal QSL, and the expanding FR1 is composed of
field lines that are anchored in the regions surrounded by the hook
of the QSL. According to the results of Titov (2007), the QSL is
rooted at the horseshoe-like features and has a helical shape
(Figures 4 and 5 in Titov 2007). The configuration of the QSL in
Titov (2007) is similar to the ``triangle-flag surface" in our
observations. The QSL in Janvier et al. (2013, 2014) is along the
J-hooked structures in the photosphere, which is also very similar
to the configuration of the hooked flare ribbon in our work. The
evolution of the fine structures of the FR1 is comparable to the
model of Janvier et al. (2013). In their simulations, the field
lines are successively formed via continuous series of reconnection
and their footpoints in one polarity move inside the QSL footprint,
following the hook shape of the QSL. The footpoints of field lines
in the other polarity are fixed, which correspond to the west fixed
footpoints in positive-polarity fields in our observations. The
slipping velocities in Janvier et al. (2013) are different from our
observations. In their model, the time-evolution of the slipping
velocities shows a high peak value that is almost 390 times as large
as the Alfv\'{e}nic speed (Figures 7 and 8 in Janvier et al. 2013).
The apparent motion of magnetic field lines is super-Alfv\'{e}nic,
and satisfies the slipping-running reconnection regime (Aulanier et
al. 2006). However, the slipping motion in our observations
satisfies the slipping reconnection regime (Aulanier et al. 2006).
Our observations have provided a 3D magnetic reconnection signal,
and more comprehensive understanding needs more observational
examples and theoretical studies.


\acknowledgments {We acknowledge the \emph{SDO}/AIA and HMI for
providing data. This work is supported by the National Basic
Research Program of China under grant 2011CB811403, the National
Natural Science Foundations of China (11303050, 11025315, 11221063
and 11003026), the CAS Project KJCX2-EW-T07 and the Strategic
Priority Research Program$-$The Emergence of Cosmological Structures
of the Chinese Academy of Sciences, Grant No. XDB09000000.}

{}
\clearpage

\begin{figure}
\centering
\includegraphics
[bb=70 115 490 709,clip,angle=0,scale=0.9]{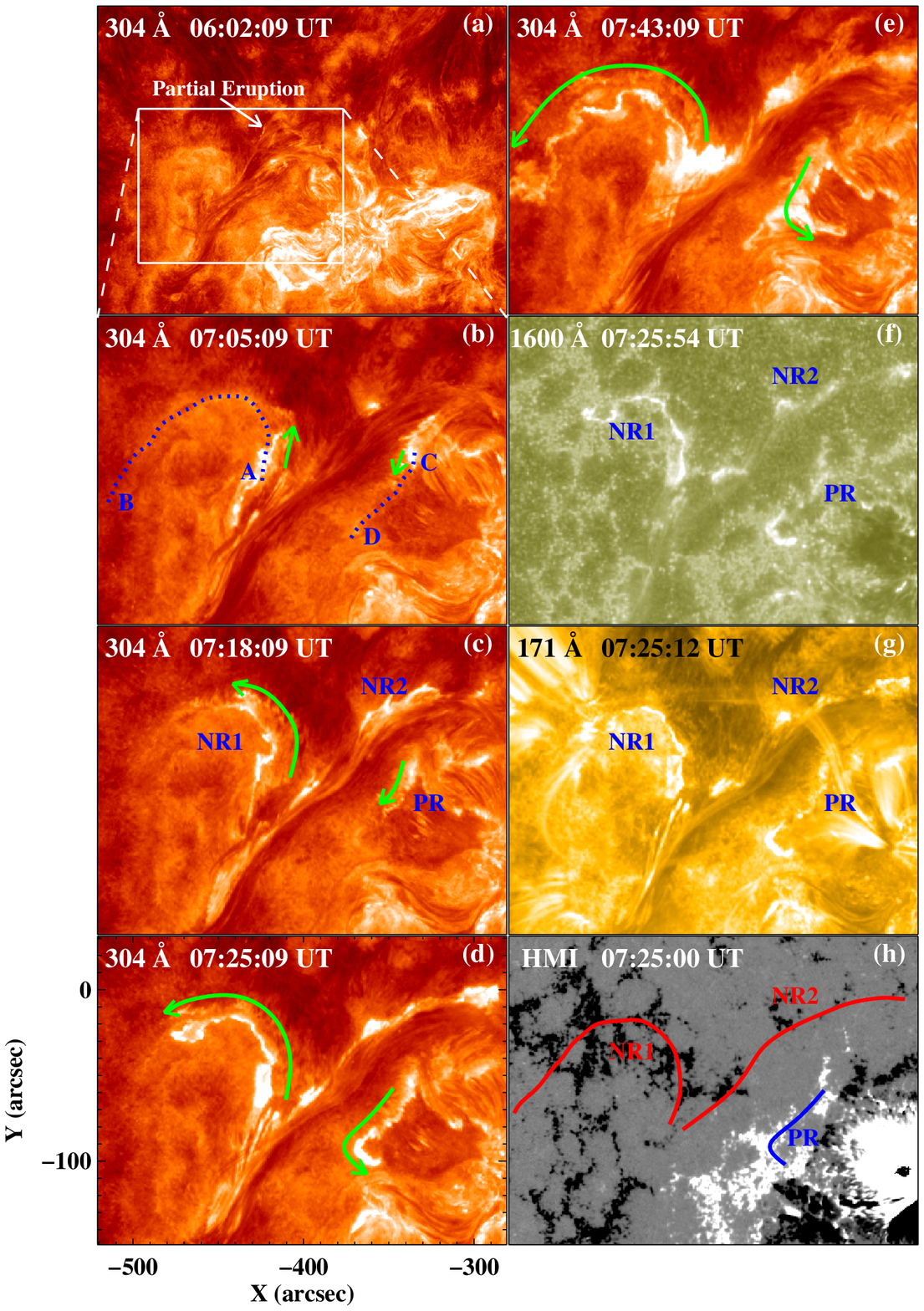}
\caption{Evolution of flare ribbons (NR1, NR2 and PR) during the
C8.9 flare on 2014 February 02 (see Animation 304). The white square
in panel (a) denotes the FOV of the images in other panels. NR1 and
NR2 are two negative-polarity ribbons and PR is one
positive-polarity ribbon (panel (h)). Green arrows point to the
opposite propagating directions of NR1 and PR. Blue dotted lines
``A$-$B" and ``C$-$D" in panel (b) show the positions of the cuts
used to obtain the stack plots shown in Figure 4. \label{fig1}}
\end{figure}
\clearpage

\begin{figure}
\centering
\includegraphics
[bb=29 143 530 677,clip,angle=0,scale=0.9]{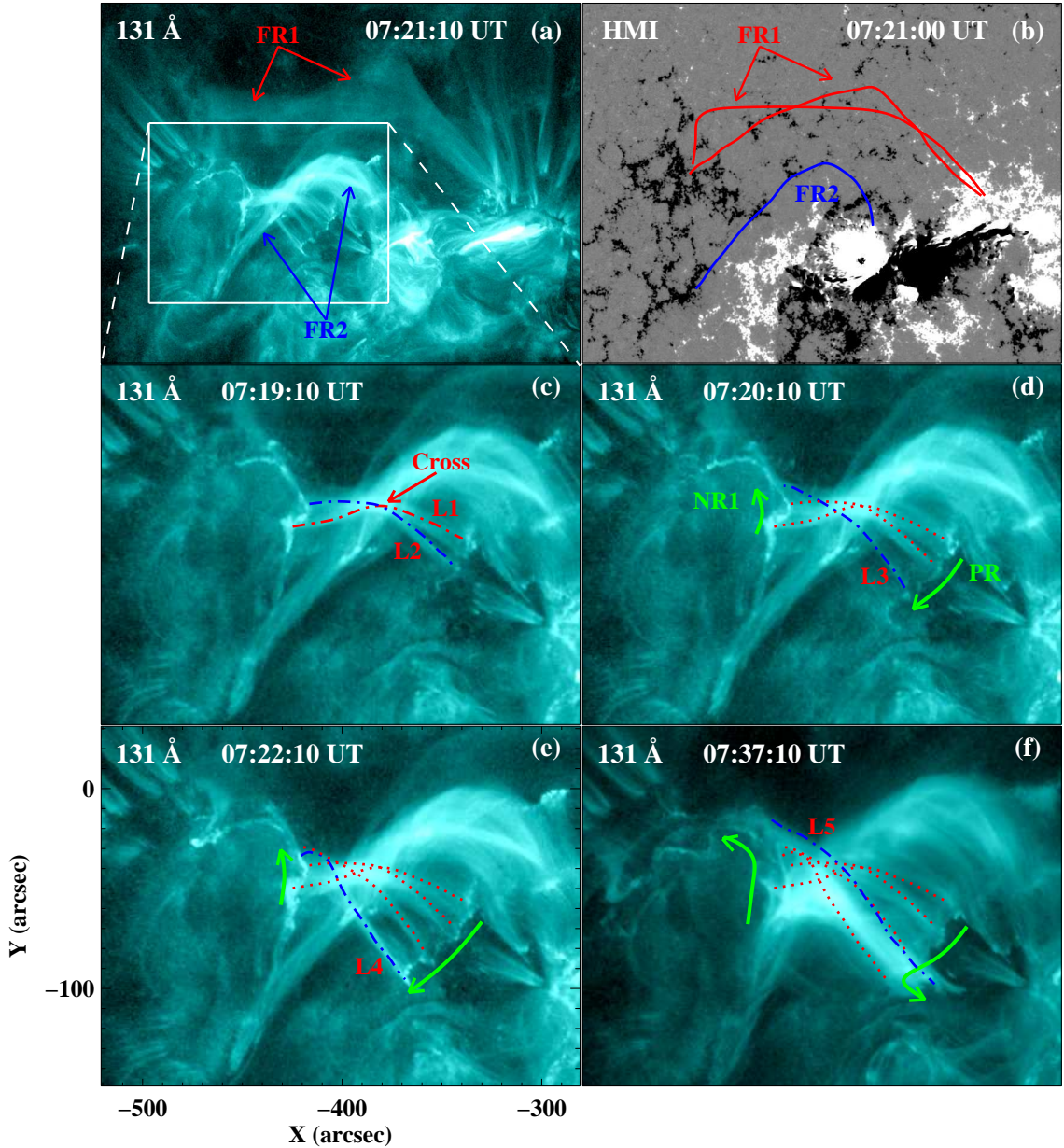}
\caption{Overview of the slipping motion of flare loops (see
Animation 131-loops). FR1 and FR2 are two flux ropes involved in
this event. The red and blue solid lines in panel (b) respectively
denote the main axes of the FR1 and FR2 at 07:21 UT. Dash-dotted
lines in panels (c)-(f) outline the flare loops connecting the NR1
with PR and dotted lines are the duplicates of the flare loops at
earlier times. Panel (f) shows the ``bi-fan" shape of flare loops.
Green arrows denote the opposite slipping directions of the loops.
\label{fig2}}
\end{figure}
\clearpage

\begin{figure}
\centering
\includegraphics
[bb=29 143 530 677,clip,angle=0,scale=0.9]{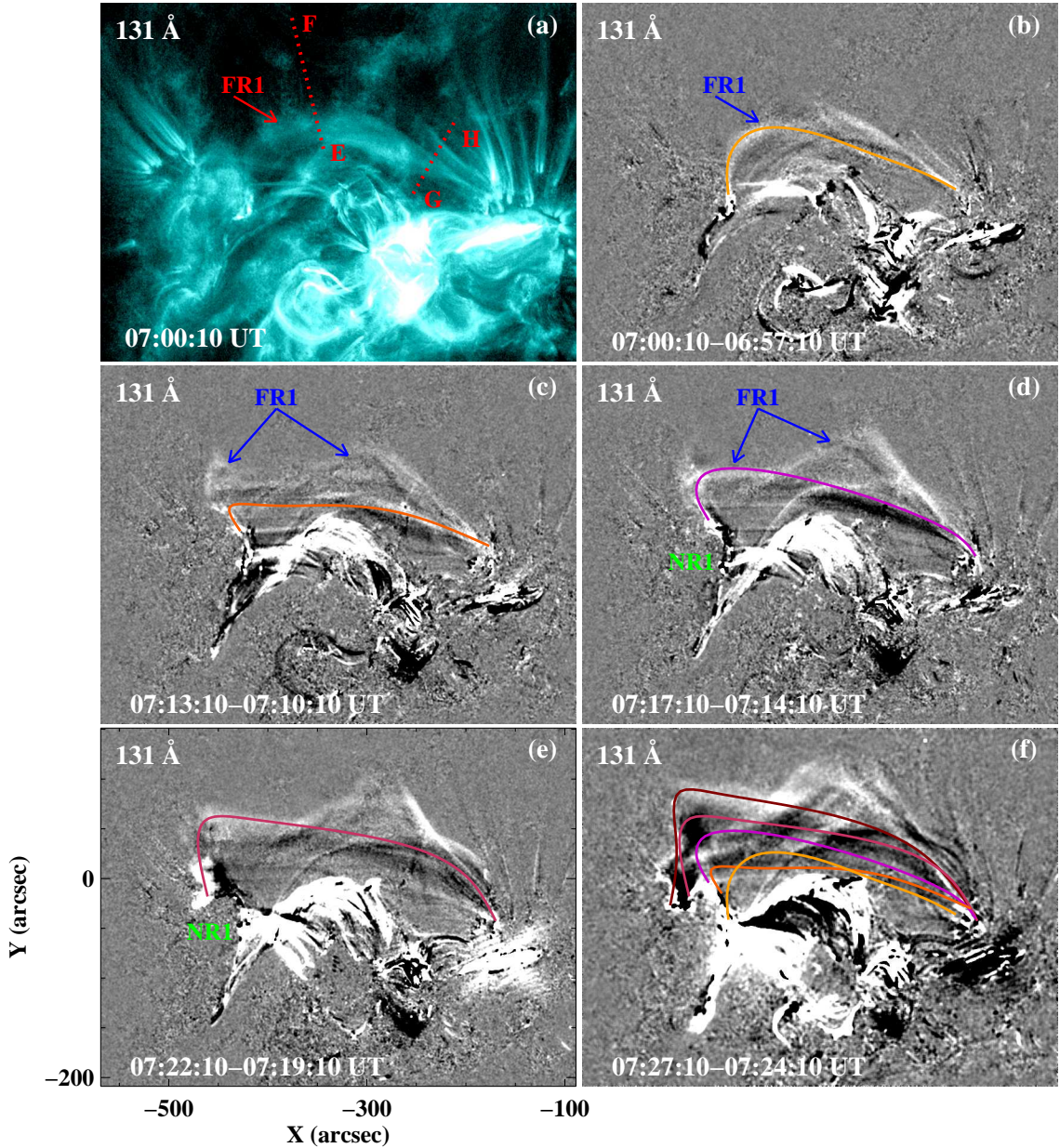}
\caption{Slipping eruption of the FR1 in AIA 131 {\AA} original and
difference images (see Animation 131-fluxropes). Red dotted curve
``E$-$F" and straight line ``G$-$H" show the positions of the cuts
used to obtain the stack plots shown in Figure 4. The solid curves
in panels (b)-(f) denote the successive appearance of the fine
structures of the FR1. Panel (f) shows the ``triangle-flag surface"
of the FR1. \label{fig3}}
\end{figure}
\clearpage

\begin{figure}
\centering
\includegraphics
[bb=100 111 476 710,clip,angle=0,scale=0.8]{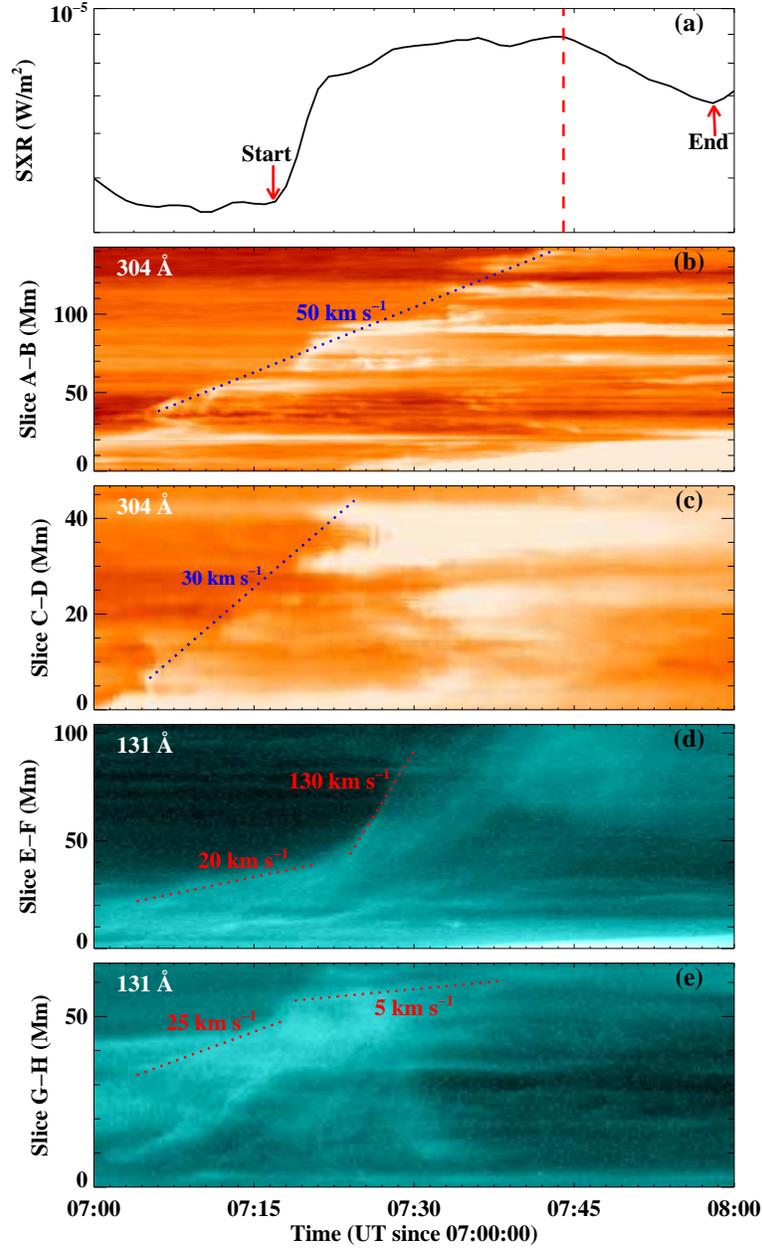} \caption{Panel
(a): GOES SXR 1$-$8 {\AA} flux of the associated flare. Red dashed
line denotes the flare peak time. Panels (b)$-$(c): stack plots
along slices ``A$-$B" and ``C$-$D" (blue dotted lines in Figure
1(b)) at 304 {\AA} respectively showing the propagations of flare
ribbons NR1 and PR. Panels (d)$-$(e): stack plots along slices
``E$-$F" and ``G$-$H" (red dotted lines in Figure 3(a)) at 131 {\AA}
respectively showing the rise of the fine structures and the whole
FR1.}
\end{figure}
\clearpage


\end{document}